\documentclass[aip,reprint]{revtex4-1}

\draft 

\usepackage{graphicx}
\usepackage{amsmath}

\begin{document}

\title{Inverse Design for Self Assembly via On-the-Fly Optimization} 

\author{Beth A. Lindquist}
\author{Ryan B. Jadrich}
\author{Thomas M. Truskett}
\email{truskett@che.utexas.edu}
\affiliation{McKetta Department of Chemical Engineering, University of Texas at Austin, Austin, Texas 78712, USA}

\date{\today}

\begin{abstract}
Inverse methods of statistical mechanics have facilitated the discovery of pair potentials that stabilize a wide variety of targeted lattices at zero temperature. However, such methods are complicated by the need to compare, within the optimization framework, the energy of the desired lattice to all possibly relevant competing structures, which are not generally known in advance. Furthermore, ground-state stability does not guarantee that the target will readily assemble from the fluid upon cooling from higher temperature. Here, we introduce a molecular dynamics simulation-based, optimization design strategy that iteratively and systematically refines the pair interaction according to the fluid and crystalline structural ensembles encountered during the assembly process. We successfully apply this probabilistic, machine-learning approach to the design of repulsive, isotropic pair potentials that assemble into honeycomb, kagome, square, rectangular, truncated square and truncated hexagonal lattices.
\end{abstract}


\maketitle 
There is a growing appreciation that a diverse array of structural motifs can be stabilized in systems of particles interacting via isotropic pair potentials, including various microphases~\cite{SALR1,SALR2,pores}, open crystalline lattices~\cite{kg_r3,AJ_PRX,ID_crystals_TS,WP_JCP}, and quasi-crystals~\cite{QC1,QC2}. Discovery of such potentials has been facilitated by inverse methods of statistical mechanics, most commonly optimization algorithms that iteratively refine the form of the interaction to attain increasingly favorable structural or thermodynamic properties.~\cite{ST_inv_des_review,AJ_inv_des_review} For cases where such strategies have been employed to find isotropic pair potentials that assemble into specific two- and three-dimensional (2D and 3D) lattices, optimization typically builds on a ground-state calculation, wherein interactions are sought that stabilize the target structure at zero temperature relative to relevant competing structures. Despite successful application of such methods to design purely repulsive interactions stabilizing multiple open lattices (including honeycomb,~\cite{AJ_PRX,ID_crystals_TS,WP_JCP} square,~\cite{AJ_PRX,ID_crystals_TS,WP_JCP} rectangular,~\cite{kg_r3} and kagome~\cite{kg_r3} in 2D and diamond,~\cite{AJ_3D,AJ_PRX,EM_dia} simple cubic,~\cite{AJ_3D,AJ_PRX} and fluorite~\cite{kg_r3} in 3D), there are some notable drawbacks to these approaches. First, they are encumbered by the requirement to specify the pool of relevant competing structures, a list that is not fully known in advance and thus must be modified as the potential is updated in the optimization. Moreover, the target structure must be checked explicitly for mechanical stability with the pair potential. Finally, interactions designed to stabilize the target structure in the ground state are not guaranteed to readily assemble the target from the fluid state upon cooling from higher temperature. 

In this Communication, we report a molecular dynamics (MD) simulation-based, inverse optimization strategy--carried out at nonzero temperature--that iteratively and systematically refines the pair interaction according to the fluid and crystalline pair structures dynamically encountered during the assembly process. The approach, while encoding practical aspects of assembly of the target from the fluid, is also technically simple and easy to implement. To illustrate the power of the methodology, we successfully employ it to design isotropic and purely repulsive pair potentials to assemble six distinct 2D lattices, including two structures which--to our knowledge--have never been stabilized via a pair potential before.

\begin{figure}[!htb]
  \includegraphics{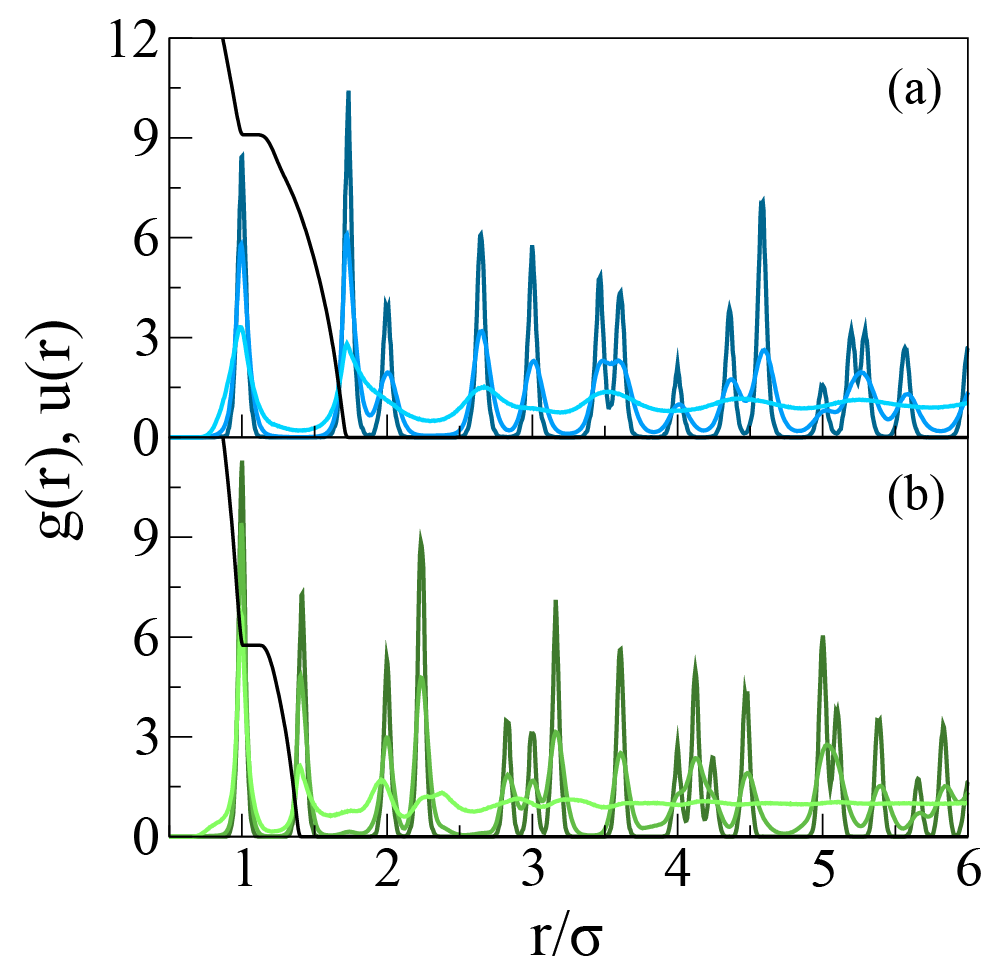}
  \caption{For the HC (a) and SQ (b) lattices, the optimized potential, $u(r)$ (thin black line), and the radial distribution functions, $g(r)$, corresponding to (from strongest to weakest structuring in the first coordination shell): the target simulation, simulation with the optimized potential, and the fluid in the final step in the optimization prior to crystallization.}
  \label{fgr:rdfs}
\end{figure} 

\begin{figure*}[!htb]
  \includegraphics{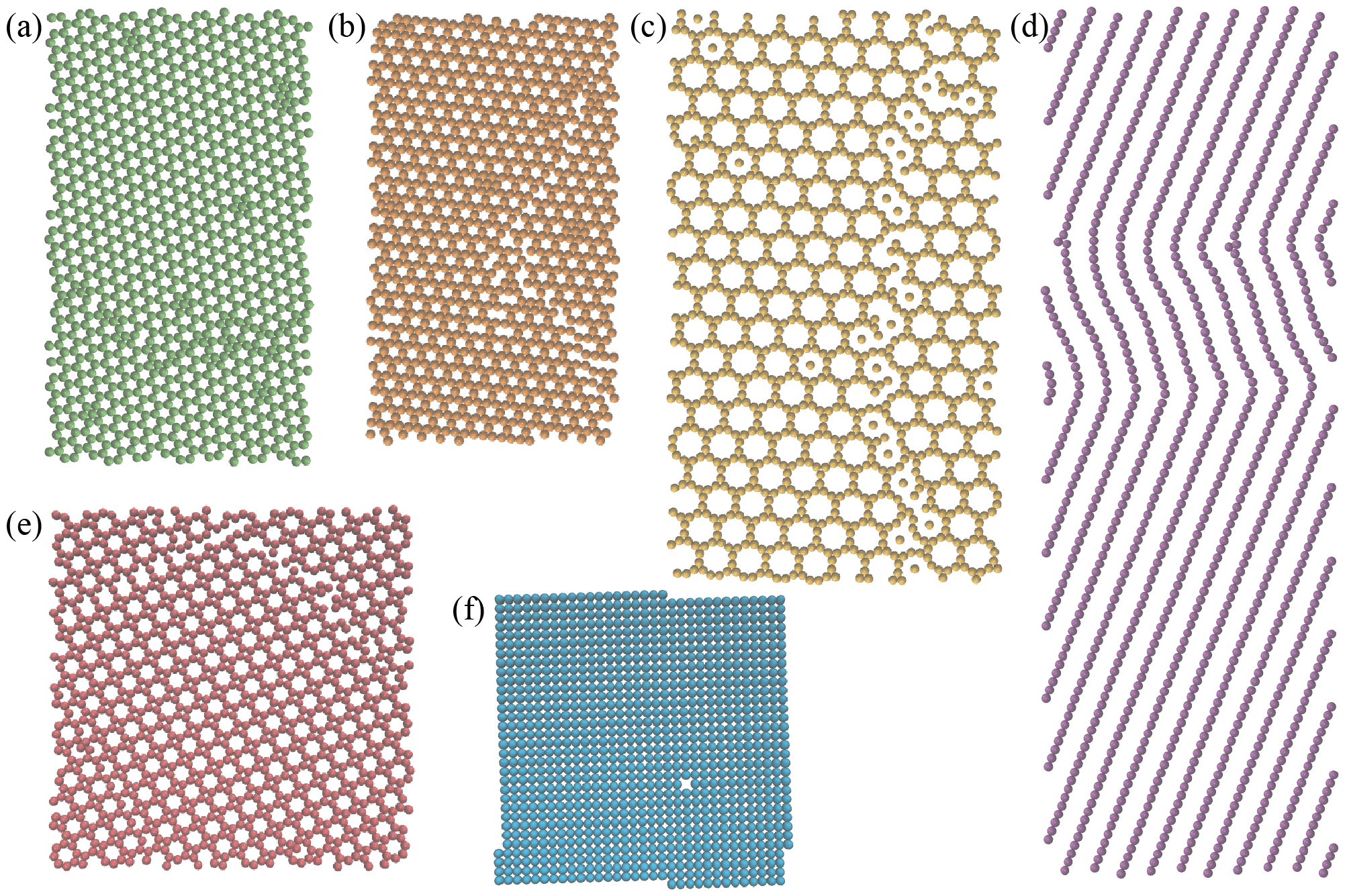}
  \caption{Final configurations after cooling to $T=0$ with $u(r)$ designed for the (a) honeycomb (HC), (b) kagome (KG), (c) truncated hexagonal (TH), (d) rectangular with an aspect ratio of three (R3), (e) truncated square (TS), and (f) square (SQ) lattices. All configurations were visualized in VMD.~\cite{VMD}}   
  \label{fgr:lattices}
\end{figure*}

Machine learning-based techniques have been successfully applied to the self-assembly of colloids, from the rational design of building blocks and templates needed to fabricate nanomaterials~\cite{design_engines,SCG_review,AAK_ML} to the elucidation of pathways involved in the assembly process.~\cite{ALF_ML} In this work, the optimization scheme we present is general and based on maximum-likelihood machine learning~\cite{barber_book} (called `relative entropy coarse-graining' in the statistical mechanics community~\cite{general_RE,IBI_and_RE_ID,RE_ID}). Within it, particle-particle interactions are tuned in order to maximize the likelihood of reproducing desired configurations (in this specific case, 2D periodic lattices). We have previously used a related optimization strategy based upon relative entropy coarse-graining to discover isotropic pair potentials that favor microphases with porous architectures of a prescribed size.~\cite{pores} 

One advantage of relative entropy coarse-graining is the expression of the pair potential $u(r|\boldsymbol{\theta})$) in terms of a functional form that is parametrized by $m$ arbitrary, scalar parameters, $\boldsymbol{\theta}\equiv[\theta_{1},\theta_{2},...,\theta_{m}]$; this allows for a variety of constraints to be straightforwardly placed on $u(r|\boldsymbol{\theta})$. For flexibility of the potential, we optimize the amplitudes of the knots for an Akima spline spaced at an interval ($\delta r$) of $\delta r / \sigma \approx0.0056$ where $\sigma$ is the nearest neighbor crystal distance. In order to avoid unphysical oscillatory potentials,~\cite{OSC_EX} we constrain the knots to be monotonically increasing with decreasing $r$ (i.e., purely repulsive interactions), though other more complex functional forms and constraints are possible. For the case of isotropic pair potentials, the updates to the parameters that characterize the potential are derived from the difference in the radial distribution functions associated with the present potential $g(r|\boldsymbol{\theta})$ and the target simulation, $g_{\text{tgt}}(r)$: 

\begin{equation} \label{eqn:update}
\begin{split}
& \boldsymbol{\theta}^{(i+1)}=\boldsymbol{\theta}^{(i)} \\
& +\alpha \int_{0}^{\infty} dr r \big[g(r| \boldsymbol{\theta}^{(i)})-g_{tgt}(r)\big] \big[\boldsymbol{\nabla}_{\boldsymbol{\theta}} u(r|\boldsymbol{\theta})\big]_{\boldsymbol{\theta}=\boldsymbol{\theta}^{(i)}} 
\end{split}
\end{equation} 
where $i$ indexes the iteration and $\alpha$ is the learning rate to be set empirically by observing the optimization stability.~\footnote{$\alpha\approx0.02$ worked well for the cases studied here.} A derivation of the update scheme and additional details pertaining to the optimization algorithm can be found in the Appendix. 

The target configurations are generated by an MD simulation where the particles are pinned to their lattice positions via a quadratic confining potential;~\footnote{The magnitude of the spring constant is chosen such that the radial distribution function has sharp crystalline features, but the peaks are still integrable. Generally this corresponded to a range of 1600-2800 $k_{B}T/\sigma^{2}$, but we do not anticipate that the exact value is critical as long as the $g(r)$ is consistent with a stable crystal and not a fluid.} all simulations contained $\geq 1000$ particles and were performed in Gromacs 4.6.5~\cite{GROMACS_1,GROMACS_2} in the NVT ensemble~\footnote{A Nos\'{e}-Hoover thermostat with a time constant of $\tau=100dt$ is employed, where $dt$ is the time step.} with periodic boundary conditions in the x and y directions. For notational convenience, we define our optimization temperature as $T^{*}$; however, the actual outcome of the optimization is $\beta u(r)$, which uniquely defines the $u(r)$ at any given temperature ($\beta=(k_{B}T)^{-1}$) that yields the desired pair structure. For the analysis below, the $u(r)$ given by $T^{*}$ is used, where $k_{B}T^{*}$ assumes a value of unity. Further details regarding simulations can be found in the Appendix. 

In contrast to prior work based on finding a potential for which the desired lattice is the ground state, we target potentials that self-assemble into the targeted lattice at finite temperature. Therefore, a key step in our procedure is to initiate each simulation from a disordered fluid state, thereby allowing all relevant structural motifs to compete as needed \emph{prior to crystallization}, including structures such as disordered microphases that might not be easily amenable to inclusion in an explicit competitor pool. Moreover, mechanical stability is directly incorporated into the optimization framework by the presence of finite temperature thermal motion. Practically, the configuration from the previous step in the optimization is melted via heating to an empirically determined temperature of $T/T^{*}=1.5$, and then subsequently cooled to $T/T^{*}=1.0$ prior to collecting statistics for the $g(r)$. We consider the optimization complete when the crystal is sufficiently stable that it no longer melts at $T/T^{*}=1.5$. 

In Fig.~\ref{fgr:rdfs}a,b, we compare the radial distribution functions of the optimized ($g(r|\boldsymbol{\theta})$) and target ($g_{\text{tgt}}(r)$) structures for both the honeycomb (HC) and square (SQ) lattice, respectively.  Both show excellent matching in the peak positions over many coordination shells, even though the ranges of the potentials ($u(r)$, in black) only span the first two coordination shells. We also show $g(r)$ from the last step before any crystallization occurred in the optimization to demonstrate that the disordered fluid locally contains muted structural signatures of the lattice. This correspondence between disordered fluid and crystalline lattice structure allows for the former to provide \emph{implicit} ``competitor pool'' information for the optimization. As the potential is optimized prior to crystallization and therefore the fluid state evolves, fluids that have structural signatures corresponding to other lattices (and therefore do not match $g_{\text{tgt}}(r)$) are suppressed (any incorrectly assembled structures are also explicitly penalized). However, because the fluid is globally disordered and highly mobile, issues such as phase boundaries and metastability that become problematic upon lattice formation do not inhibit proper sampling of phase space in the fluid.

\begin{figure}[!htb]
  \includegraphics{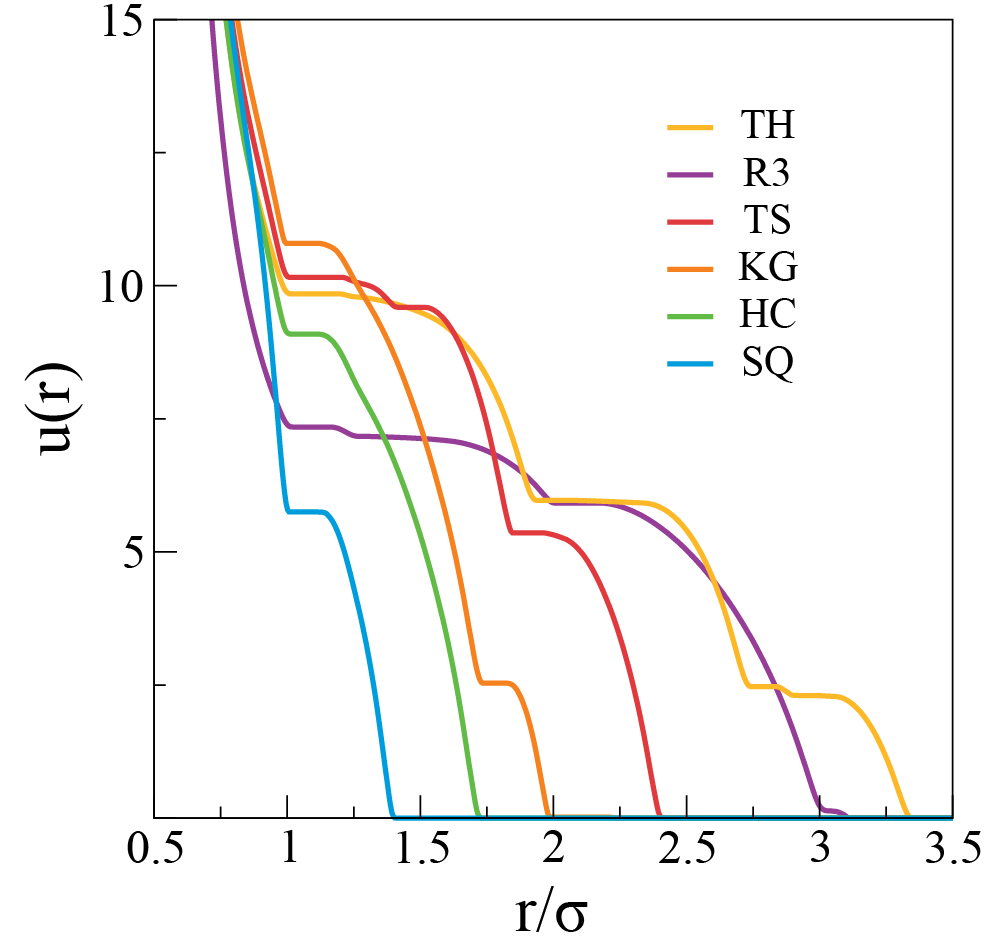}
  \caption{Optimized pair potentials for the lattices considered in this work. The figure legend is ordered by decreasing range in $u(r)$ from top to bottom.}
  \label{fgr:potentials}
\end{figure}

In addition to HC and SQ, this optimization procedure was carried out for the rectangular lattice with an aspect ratio of three (R3), kagome (KG), truncated square (TS), and truncated hexagonal (TH) lattices. The resulting potentials were then simulated shortly at a sufficiently high temperature to fully melt the crystal, and then the simulation was very slowly cooled through the empirically determined melting-freezing transition range and then further cooled to $T=0$. The resulting structures are shown in Fig ~\ref{fgr:lattices}a-f, where it is clear that the optimizations were successful. Only small expected defects are present due to finite size of the periodically replicated simulation cell and any misalignment of the nucleated crystal with the cell. The corresponding potentials are shown in Fig.~\ref{fgr:potentials}. The potentials are essentially repulsive shoulders, with one to three such features beyond the core. Shoulders originate from the monotonicity constraint, which prevents the development of attractive wells at specific coordination shells. Instead, $u(r)$ develops features with stiff repulsive forces 1) to penalize the surmounting of the shoulder, while 2) yielding a strong ``thermodynamic'' pressure (i.e., the ensemble averaged force due to the influence of every other particle) that pushes two particles into the shoulder. In essence, attractive wells are forgone for stiff repulsions and higher pressures to build the strong, specific coordination needed for targeting a crystal phase.      

\begin{figure}[!htb]
  \includegraphics{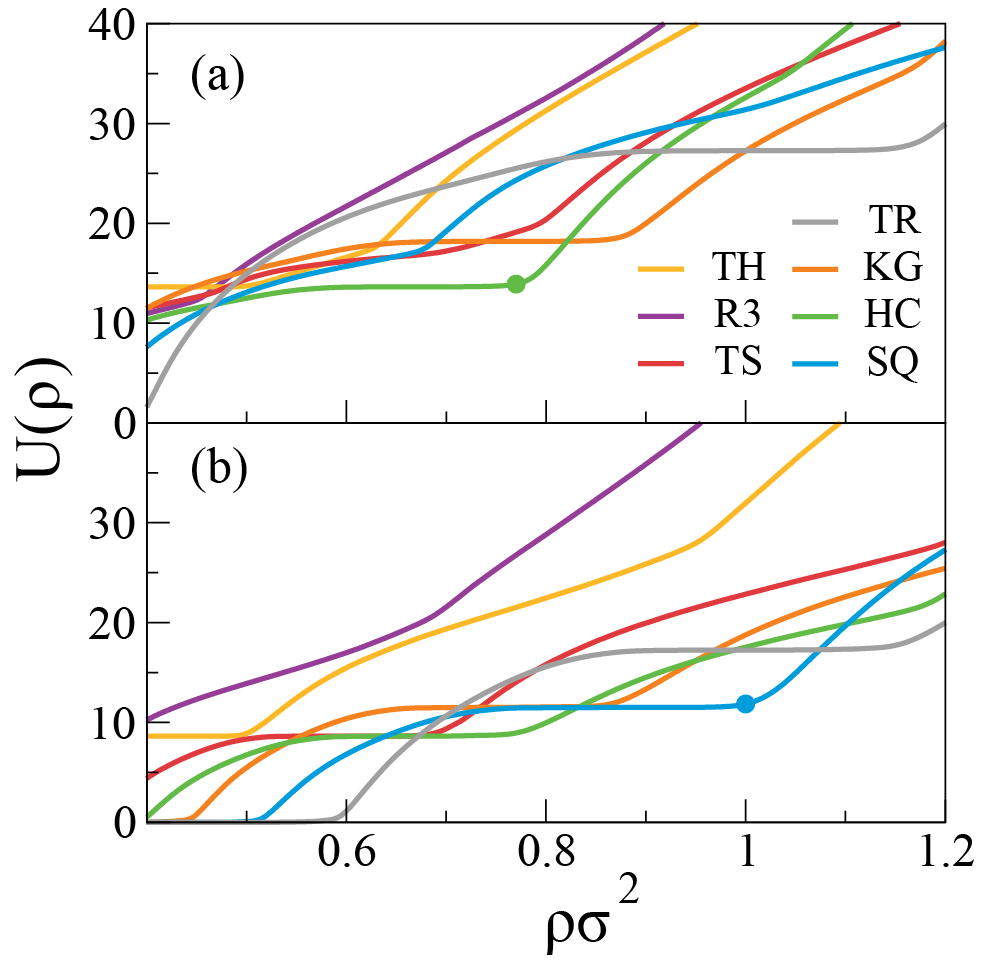}
  \caption{Ground state energies for the lattices optimized in this work, plus a triangular lattice, for (a) $u_{\text{HC}}(r)$ and (b) $u_{\text{SQ}}(r)$ as a function of $\rho$. The circle denotes the density where the optimization was performed, $\rho_{\text{opt}}$. At $\rho_{\text{opt}}$, the potentials are ordered as (a) HC, KG, TS, SQ, TR, TH, R3 and (b) SQ, TR, HC, KG, TS, TH, R3 with increasing energy.}
  \label{fgr:gs}
\end{figure}

While the optimization framework presented here does not guarantee that the assembled crystal is also the ground state, we can confirm that the desired lattice, where the particles are in their ideal positions, is the lowest energy state at $T=0$ of the lattices featured in this work. In Fig.~\ref{fgr:gs}a,b, we show the total energy, $U$, as a function of $\rho$ for the six lattices studied here in addition to the triangular (TR) lattice, using $u_{\text{HC}}(r)$ and $u_{\text{SQ}}(r)$, respectively. The circle denotes the optimization density for the lattice, and we see that the desired lattice is the lowest in energy for a reasonable range about $\rho$ in the optimization, though the values of $\rho$ outside of the optimization point are not guaranteed to be mechanically stable. The other potentials display similar behavior (data not shown). 

\begin{figure}[!htb]
  \includegraphics{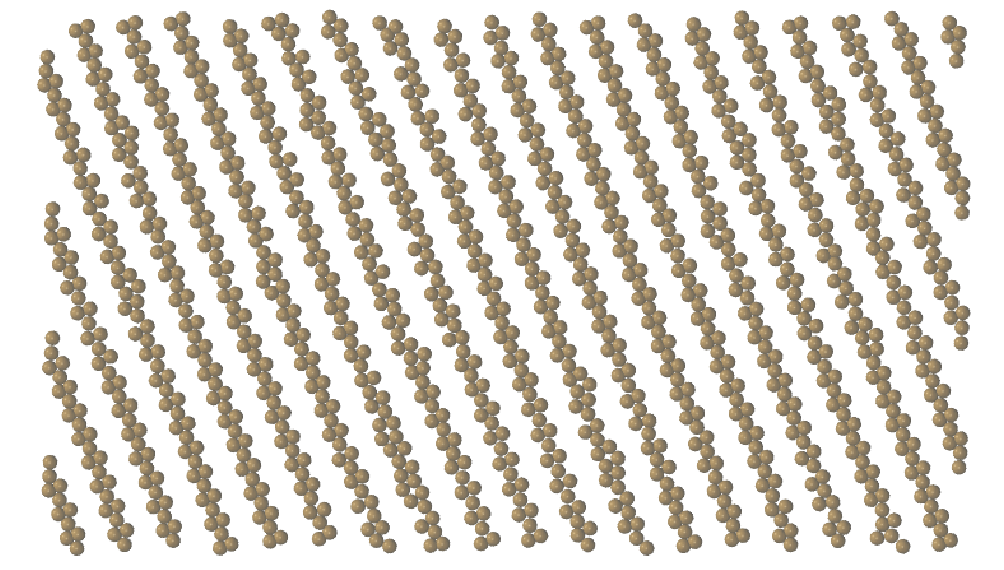}
  \caption{Self-assembled structure of $u(r)$ discovered with an optimization targeting the TH lattice, but performed without starting each step from a fluid configuration.}
  \label{fgr:stripes}
\end{figure}

The importance of heating the system into the fluid phase at each optimization step, thereby incorporating the assembly process into the optimization, can be demonstrated in the context of the TH lattice. If optimizations are performed beginning from the crystalline state, then mechanical stability is accounted for since the crystal can fall apart in the MD simulation, but the role of competitor phases to the self-assembly process \emph{is not included}. Simulated annealing of a potential optimized for the TH lattice without melting the crystal at the outset of each iteration resulted in assembly of a crystalline stripe phase as shown in Fig.~\ref{fgr:stripes}. The energies at $T=0$ of the resulting (defective) striped phase and the perfect TH lattice are similar, with the latter being more stable by ~5\%. However, the striped phase forms first and is sufficiently kinetically stable that it never transitions into the TH lattice, even with very slow annealing schedules. Therefore, the TH phase appears to be kinetically inaccessible via simulation, though it might in fact be the ground state. Beginning with a disordered state in every optimization step circumvents this difficulty entirely because the $g(r)$ is collected from the assembled phase, and the update scheme therefore adjusts the potential accordingly if an incorrect structure is encountered.

In summary, we have introduced a simple, relative entropy based inverse design approach, which we demonstrated can successfully discover purely repulsive, isotropic pair potentials that favor assembly of particles into a wide variety of open 2D lattices. Because this method is based on standard, MD simulation techniques, there is no need to construct or update a large competitor pool during the optimization. Instead, structural information is simply encountered in the simulation and utilized in the optimization on-the-fly, via self-assembly, to target a given crystal--including states that would otherwise be hard to predict \emph{a priori} or to incorporate into a competitor pool (such as periodic microphases). In addition to the major simplification from the implicit competitor pool, mechanical stability, and likewise some finite degree of thermal stability, are naturally encoded in the method. However in this approach, it is not known whether the desired lattice is the ground state; rather, the optimization gives insight into what structure will result from self-assembly at nonzero temperature for a given potential. Also, it should be noted that crystals with large kinetic barriers to self-assembly will likely be difficult to treat with this strategy, though it is also reasonable to suspect that such lattices will encounter real-world complications that may preclude their assembly in general--particularly for large nanoparticle to micron-sized colloidal systems. 

Intriguing avenues for future work include optimizing other constrained functional forms for $u(r)$, and the use of different types of target structures, such as 3D lattices or binary lattices with multiple interactions. With respect to the former, one strategy to move towards realization of such assemblies would be to constrain $u(r|\boldsymbol{\theta})$ to experimentally motivated models that describe, for example, ligand-coated colloids.~\cite{CollInt1,CollInt2} Some existing interaction models, for micelles~\cite{QC1} for instance, bear resemblance to the simpler potentials presented in this work (HC and SQ in particular). However, simplification of the remaining potentials via greater constraint on the functional form might be necessary to realize the corresponding lattices.

\begin{acknowledgments}
T.M.T. acknowledges support of the Welch Foundation (F-1696) and the National Science Foundation (CBET-1403768). We also acknowledge the Texas Advanced Computing Center (TACC) at the University of Texas at Austin for providing computing resources used to obtain results presented in this paper.

\end{acknowledgments}

\setcounter{equation}{0}
\renewcommand\thetable{A\arabic{table}}
\renewcommand{\thesection}{\thepart .\arabic{section}}
\renewcommand\theequation{A\arabic{equation}}

\section*{Appendix}

\subsection{Derivation of Update Scheme}

Here we provide a brief derivation of the update scheme shown in Eqn. 1 of the main text. 

The probability of observing a configuration, $\textbf{R}_{i}$, in the canonical ensemble is given by the Boltzmann factor normalized by the partition function, or $Z$. Therefore the probability of observing $M$ statistically independent and identically distributed configurations, $\textbf{R}_{1:M}$, in the NVT ensemble is given by the product of such terms:

\begin{equation} \label{eqn:likelihood}
P(\textbf{R}_{1:M}|\boldsymbol{\theta}) \equiv \prod_{i=1}^{M} P(\textbf{R}_{i}|\boldsymbol{\theta}) = \prod_{i=1}^{M} \dfrac{\exp[-\beta U(\textbf{R}_{i}|\boldsymbol{\theta})]}{Z(\boldsymbol{\theta})}
\end{equation} 
where $U(\textbf{R}_{i}|\boldsymbol{\theta})$ is the potential energy for configuration $\textbf{R}_{i}$ and $\boldsymbol{\theta}$ is a vector of the tunable values that parametrize the potential. (This quantity is also termed \emph{the likelihood} of $\boldsymbol{\theta}$ given $\textbf{R}_{1:M}$.)

However, we need to optimize $\boldsymbol{\theta}$ for a given set of configurations sampled from the target simulation (as described in the main text). Therefore, we are actually interested in the quantity $P(\boldsymbol{\theta}|\textbf{R}_{1:M})$, i.e., the probability of the parameters $\boldsymbol{\theta}$ given the configurations $\textbf{R}_{1:M}$. Within Bayesian statistics, this quantity is termed the posterior distribution and depends on $P(\textbf{R}_{1:M}|\boldsymbol{\theta})$ as follows:
\begin{equation} \label{eqn:posterior}
P(\boldsymbol{\theta}|\textbf{R}_{1:M}) \equiv \dfrac{P(\textbf{R}_{1:M}|\boldsymbol{\theta})P(\boldsymbol{\theta})}{P(\textbf{R}_{1:M})}
\end{equation} 
In Eqn.~\ref{eqn:posterior}, $P(\boldsymbol{\theta})$ and $P(\textbf{R}_{1:M})$ are marginalized probabilities, or the probabilities in the absence of any information about the other variable. Therefore, $P(\boldsymbol{\theta})$ has the interpretation of a ``prior'' distribution--reflecting any prior knowledge pertaining to what the parameters should be--and $P(\boldsymbol{\theta}|\textbf{R}_{1:M})$ reflects how this prior is updated once observational information ($\textbf{R}_{1:M}$) is taken into account. The latter is the function that we seek to maximize with respect to $\boldsymbol{\theta}$:
\begin{equation} \label{eqn:argmax_posterior}
\text{argmax}_{\boldsymbol{\theta}}P(\boldsymbol{\theta}|\textbf{R}_{1:M})
\end{equation}

In our current work, we assume a uniform distribution for $\boldsymbol{\theta}$. With a flat prior for $\boldsymbol{\theta}$, and because $P(\textbf{R}_{1:M})$ is $\boldsymbol{\theta}$-independent, it is easy to see that $\text{argmax}_{\boldsymbol{\theta}}P(\boldsymbol{\theta}|\textbf{R}_{1:M}) = \text{argmax}_{\boldsymbol{\theta}}P(\textbf{R}_{1:M}|\boldsymbol{\theta})$. Thus, for the remainder of this text we seek to maximize the likelihood $P(\textbf{R}_{1:M}|\boldsymbol{\theta})$ with respect to $\boldsymbol{\theta}$, and maximization of the posterior distribution under the above assumptions is referred to as the maximum likelihood approach. 

In practice, it is easier to maximize the log-likelihood. Taking the natural log of Eqn.~\ref{eqn:likelihood} and dividing by $M$ yields:
\begin{equation} \label{eqn:log_likelihood}
\begin{split}
& \dfrac{1}{M}\text{ln}P(\textbf{R}_{1:M}|\boldsymbol{\theta}) \equiv \dfrac{1}{M}\sum_{i=1}^{M}\text{ln} P(\textbf{R}_{i}|\boldsymbol{\theta}) \\
& = -\dfrac{1}{M}\sum_{i=1}^{M}\Big[\beta U(\textbf{R}_{i}|\boldsymbol{\theta})\Big] - \text{ln}Z(\boldsymbol{\theta})
\end{split}
\end{equation}
which can be written as 
\begin{equation} \label{eqn:log_likelihood_2}
\langle\text{ln}P(\textbf{R}|\boldsymbol{\theta})\rangle_{P_{\text{tgt}}(\textbf{R})} = -\langle\beta U(\textbf{R}|\boldsymbol{\theta})\rangle_{P_{\text{tgt}}(\textbf{R})} - \text{ln}Z(\boldsymbol{\theta})
\end{equation}
in the large configuration limit, i.e., $M \rightarrow \infty$. $P_{\text{tgt}}(\textbf{R})$ is the probability distribution of the target simulation (from which the configurations are actually sampled).

In order to carry out the log-likelihood maximization, we employ a gradient ascent optimization algorithm, 
\begin{equation} \label{eqn:grad_descent}
\boldsymbol{\theta}^{(i+1)} = \boldsymbol{\theta}^{(i)} + \alpha \big[\boldsymbol{\nabla}_{\boldsymbol{\theta}} \langle\text{ln}P(\textbf{R}|\boldsymbol{\theta})\rangle_{P_{\text{tgt}}(\textbf{R})}\big]_{\boldsymbol{\theta}=\boldsymbol{\theta}^{(i)}}
\end{equation}
where $\alpha$ is the empirically determined step size. From Eqn.~\ref{eqn:log_likelihood_2} and employing the relation $Z(\boldsymbol{\theta})\equiv\int d\textbf{R} \exp[-\beta U(\textbf{R}|\boldsymbol{\theta})]$, we find that 

\begin{equation} \label{eqn:grad_log_likelihood}
\begin{split}
& \boldsymbol{\nabla}_{\boldsymbol{\theta}} \langle\text{ln}P(\textbf{R}|\boldsymbol{\theta})\rangle_{P_{\text{tgt}}(\textbf{R})} = \\
& -\langle \boldsymbol{\nabla}_{\boldsymbol{\theta}} \beta U(\textbf{R}|\boldsymbol{\theta})\rangle_{P_{\text{tgt}}(\textbf{R})} + \langle \boldsymbol{\nabla}_{\boldsymbol{\theta}} \beta U(\textbf{R}|\boldsymbol{\theta})\rangle_{P(\textbf{R}|\boldsymbol{\theta})} 
\end{split}
\end{equation}

For the specific case of an isotropic pair potential, i.e., $U(\textbf{R}|\boldsymbol{\theta})\equiv \dfrac{1}{2}\sum_{i \neq j}^{N}u(r_{i,j}|\boldsymbol{\theta})$, we can factorize the terms on the righthand side of Eqn.~\ref{eqn:grad_log_likelihood} into integrals over the product of the gradient with respect to $\boldsymbol{\theta}$ of the pair potential, $\boldsymbol{\nabla}_{\boldsymbol{\theta}} u(r|\boldsymbol{\theta})$, and an averaged two-point density. Because the potential only depends on pair interactions, only a two-point correlation function is required. Therefore, in two dimensions, the gradient required by Eqn.~\ref{eqn:grad_descent} can be rewritten as 
\begin{equation} \label{eqn:grad_log_likelihood_2}
\begin{split}
& \boldsymbol{\nabla}_{\boldsymbol{\theta}} \langle\text{ln}P(\textbf{R}|\boldsymbol{\theta})\rangle_{P_{\text{tgt}}(\textbf{R})} = \\
& \pi \rho N \int_{0}^{\infty} drr[g(r|\boldsymbol{\theta})-g_{\text{tgt}}(r)]\boldsymbol{\nabla}_{\boldsymbol{\theta}}u(r|\boldsymbol{\theta})
\end{split}
\end{equation}
after employing the definition of the radial distribution function, $g(r)$, performing one of the integrals over all space, and converting to spherical coordinates. Inserting this term into Eqn.~\ref{eqn:grad_descent} and absorbing all constants into $\alpha$ yields Eqn. 1 in the main text.

Finally, note that the lefthand side of Eqn.~\ref{eqn:log_likelihood_2} differs from the Kullback-Leibler divergence in the large sample limit, $\big<\text{ln}P_{\text{tgt}}(\textbf{R})\big>_{P_{\text{tgt}}(\textbf{R})}-\big<\text{ln}P(\textbf{R}|\boldsymbol{\theta})\big>_{P_{\text{tgt}}(\textbf{R})}$, only by a constant in $\boldsymbol{\theta}$ and in sign. Therefore minimization of the Kullback-Leibler divergence, a typical starting point for the derivation of relative entropy coarse-graining procedures, is equivalent to Eqn. 1 in the main text. 

\subsection{Additional Optimization and Simulation Details}

The number of iterations needed in the optimization varied with lattice type, but generally approximately 100-200 iterations were required. Therefore, for the iterative optimization scheme, relatively short simulations of $8\text{x}10^{6}$ steps were needed for computational efficiency of the optimization. The first half of these simulations entailed cooling from $T/T^{*}=1.5$ to $T/T^{*}=1.0$, where $T^{*}$ is defined as the optimization temperature, yielding a cooling rate of $1.25\text{x}10^{-7}T^{*}/dt$. (The time step for the MD simulations is also defined in terms of $T^{*}$: $dt/\sqrt{\sigma^{2}m\beta^{*}} \approx 0.001$, where $\beta^{*}=(k_{B}T^{*})^{-1}$.) The second half of the simulation was run at $T^{*}$; 667 configurations evenly distributed over the final $1\text{x}10^{6}$ steps were used to collect statistics to compute $g(r|\boldsymbol{\theta})$.

Subsequent simulations with the optimized potentials were much longer, $1\text{x}10^{8}$ steps, $5\text{x}10^{7}$ of which were spent cooling between the melting and freezing temperatures as described in the main text. The cooling rate depended on the gap in temperature between these two end points, varying between $8.3\text{x}10^{-10}T^{*}/dt$ and $2.5\text{x}10^{-9}T^{*}/dt$ for the different lattices. We observed no discernible dependence of the self-assembly towards slowing the cooling rate over orders of magnitude. Below the freezing temperature (i.e., after large-scale assembly has already occurred), the annealing schedule was accelerated to $\sim 2\text{x}10^{-8}T^{*}/dt$ as generally only small-scale local rearrangements occurred in this regime. 

\begin{table}[h!]
\centering
 \begin{tabular}{|c || c | c | c |} 
 \hline
  & $r_{n}/\sigma$ & $x/\sigma$ & $y/\sigma$ \\ 
 \hline
 HC & 1.74 & 27.7 & 48.0 \\ 
 SQ & 1.42 & 32.0 & 32.0 \\
 KG & 2.24 & 45.0 & 26.0 \\
 R3 & 3.11 & 32.0 & 96.0 \\
 TS & 2.42 & 41.0 & 41.0 \\ 
 TH & 3.35 & 37.4 & 64.7 \\ 
 \hline
 \end{tabular}
 \caption{Cut-offs for $u(r)$ and simulation box sizes for each lattice.}
\end{table}

The cut-off for $u(r)$ was systematically chosen to include as few coordination shells as possible while still allowing for the correct lattice to form. The resulting cut-offs are given in Table A1, in addition to the box lengths in the $x$ and $y$ directions for the optimization and subsequent simulations.

If the position of the knots in the Akima spline are defined as the vector $\boldsymbol{r}=$\{$r_{1}, r_{2}, ..., r_{n}$\}, where $r_{n}$ is the cut-off for the potential, then each knot has a corresponding tunable parameter $\theta$, i.e., $\boldsymbol{\theta} \equiv \theta(\boldsymbol{r})$. In order to simply enforce the monotonicity constraint, the $\theta(\boldsymbol{r})$ values do not directly encode the knot amplitudes ($\boldsymbol{\kappa}$) but rather indicate the difference between the amplitude of the present knot and the next knot:

\begin{equation}
\kappa(r_{i}) = \kappa(r_{i+1}) + \theta(r_{i})
\end{equation}
where $\kappa(r_{n})$ is fixed at zero. As a result, enforcing the repulsive constraint is equivalent to requiring that all $\theta$ values are non-negative.

The starting guess for the optimization is a power law weighted by step-like smoothing function to minimize the force near and at that cutoff
\begin{equation}
\beta u_{0}(r) \equiv \dfrac{1}{2} A\bigg(\dfrac{\sigma}{r}\bigg)^{a} \bigg(1-\tanh \bigg[k\bigg(\dfrac{r}{\sigma} - \dfrac{r_{s}}{\sigma}\bigg)\bigg]\bigg)
\end{equation}
where $A$ is the dimensionless amplitude, $k$ controls the steepness of the switching function, $a$ is the power, $r_{\text{s}}$ is the center of the smoothing function, and $\sigma$ is the nearest neighbor crystal distance. In this publication we set $A\equiv1.8$, $k\equiv8.9$,  $a\equiv5$ and $r_{\text{s}}$ is set near to, but before, the potential cut-off. However, we have not found the optimization to be sensitive to the details of the initial guess. Moreover, Ref. 20 in the main text notes that for linear parameters, such as $\boldsymbol{\theta}$ as defined above, there is only one maximum on the landscape. 

\section*{References}

%

\end{document}